\documentclass[12pt,preprint,referee]{aastex}

\usepackage[usenames]{color}

\usepackage{graphics}
\usepackage{graphicx}
\usepackage{epstopdf}
\usepackage{url}
\usepackage[stable]{footmisc}

\newcommand{\be}{\begin{equation}}
\newcommand{\ee}{\end{equation}}
\newcommand{\ba}{\begin{eqnarray}}
\newcommand{\ea}{\end{eqnarray}}
\newcommand{\bc}{\begin{center}}
\newcommand{\ec}{\end{center}}
\newcommand{\lsi}{LS~I~+61$^{\circ}$303 }
\newcommand{\ls}{LS 5039}
\newcommand{\fg}{1FGL J1018.6-5856}

\begin{document}

\title{\textsc{INTEGRAL observations of the $\gamma$-ray binary 1FGL J1018.6-5856 }}
\author{Jian Li\altaffilmark{1}, Diego F. Torres\altaffilmark{2,3}, 
Yupeng Chen\altaffilmark{1}, Diego G\"{o}tz\altaffilmark{4},\\
Nanda Rea\altaffilmark{3}, Shu Zhang\altaffilmark{1}, 
G. Andrea Caliandro\altaffilmark{3}, Jianmin Wang\altaffilmark{1,5}}

\altaffiltext{1}{Laboratory for Particle Astrophysics, Institute of High
Energy Physics, Beijing 100049, China. }
\altaffiltext{2}{Instituci\'o Catalana de Recerca i Estudis Avan\c{c}ats (ICREA).}
\altaffiltext{3}{Institut de Ci\`encies de l'Espai (IEEC-CSIC),
              Campus UAB,  Torre C5, 2a planta,
              08193 Barcelona, Spain}
\altaffiltext{4}{AIM (UMR 7158 CEA/DSM-CNRS-Universit\'e Paris Diderot) Irfu/Service d'Astrophysique,  Saclay, 91191 Gif-sur-Yvette Cedex,  France}

\altaffiltext{5}{Theoretical Physics Center for Science Facilities (TPCSF), CAS, China}

\begin{abstract}

The {\it Fermi}-LAT collaboration has recently reported that one of their detected sources, namely, 1FGL J1018.6-5856,
is a new gamma-ray binary similar to \ls. This has prompted
efforts to study its multi-frequency behavior. In this report, 
we present the results from   5.78-Ms  \textit{INTEGRAL} 
IBIS/ISGRI observations on the source 1FGL J1018.6-5856. 

By combining all the available \textit{INTEGRAL} data, a detection is made at a significance level of 5.4$\sigma$ in the 18--40 keV band, with an average intensity of 0.074 counts s$^{-1}$. 
However,  we find that, there is non-statistical noise in the image that effectively reduces the significance to about 4$\sigma$ and a significant part of the signal appears to be located in a 0.2-wide phase region, at phases 0.4--0.6 (where even the corrected significance amounts to 90\% of the total signal found). Given the scarcity of counts, a variability is hinted at about $3\sigma$ at the hard X-rays, with an anti-correlation  with the {\it Fermi}-LAT periodicity. Should this behavior be true,
it would be similar to that found in \ls, and prompt observations with TeV telescopes at phases anti-correlated with the GeV maximum.

\end{abstract}

\keywords{X-rays: binaries, X-rays: individual (1FGL J1018.6-5856)} 

\section{Introduction}

The Fermi Large Area telescope ({\it Fermi}-LAT) is a 
pair-production detector with large effective area ($\sim$ 8000 cm$^2$ on axis for $E >1$ GeV) and field of view ($\sim$ 2.4 sr at 1 GeV), sensitive to gamma rays in the energy range from 20 MeV to $>$ 300 GeV (Atwood et al. 2009). Precisely because of its capability of covering a wide region of the sky, its normal mode of operation is surveying, which facilitates serendipitous discoveries and simultaneous observations of many sources. 
The {\it Fermi}-LAT collaboration has recently released (Corbet et al. 2011) the results for continuing data analysis of the gamma-ray emission from 1FGL J1018.6-5856, for which earlier information was also reported in the {\it Fermi}-LAT source catalog (Abdo et al., 2010). The considered data,
obtained between MJD 54682 and 55627 in the energy range 100 MeV to 200 GeV show the presence of periodic modulation with a period of 16.58 $\pm$ 0.04 days, and an epoch of maximum gamma-ray flux at MJD 55403.3 $\pm$ 0.4
(Corbet et al. 2011).
A coincident X-ray flux was found using {\it Swift}-XRT observations, which features also a high degree of 
variability, with the 0.3--10 keV count rates ranging from approximately 0.01 to 0.05 counts/s, as well as a star of magnitude B2
which in turn coincides with the  {\it Swift}-XRT detection. Based on all of the former, Corbet et al. (2011) reported that \fg\
is a new gamma-ray binary like, for instance, LS 5039 (see Abdo et al. 2009).
In this Letter we present the results of the analysis of  5.78-Ms   \textit{INTEGRAL} IBIS/ISGRI data on the source 1FGL J1018.6-5856.

\section{Observations and data analysis}

\textit{INTEGRAL} (Winkler et al. 2003) is optimized to work between 15 keV--10 MeV. Its main instruments are the IBIS (15 keV--10 MeV; Ubertini et al. 2003) and the Spectrometer on board \textit{INTEGRAL}  (SPI, 20 keV--8 MeV; Vedrenne et al. 2003). At the lower energies
(15 keV--1 MeV), the cadmium telluride array ISGRI (Lebrun et al.
2003) of IBIS has a better continuum sensitivity than SPI below $\sim 300$ keV.
%Because of the coded-mask design of the instruments, the satellite normally operates in dithering mode, which suppressesthe systematic effects on spatial and temporal backgrounds.
 The \textit{INTEGRAL} observations were carried out per pointing, called as individual Science Windows (SCWs), with a typical time duration of about
2000 s each. The data reduction was performed by using the
standard Online Science Analysis (OSA), version 9.0. The results
were obtained by running the pipeline from the raw to the image level. 
In this analysis, only IBIS/ISGRI public data are taken into account. The available INTEGRAL observations when 1FGL J1018.6-5856 had offset angle less than 14$^o$ comprise about  2014 SCWs, adding up to a total exposure time of 5.78 Ms. Our dataset covers rev. 30--867, from 2003-01-11 to 2009-11-20 (MJD 52650-55155). This large amount of data allow for an in-depth investigation on the hard X-ray emission from 1FGL J1018.6-5856.

\section{Results}

\begin{figure*}[t]
\centering
\includegraphics[angle=0, scale=0.5] {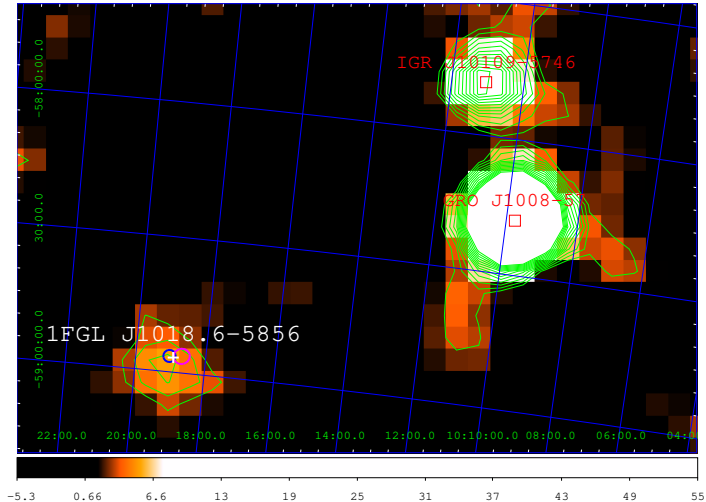}
\caption{ Mosaic image of the \fg\ sky region, derived by
combining all ISGRI data in the 18--40 keV band. The strongest source is
GRO J1008-57 whereas the faintest one, visible only in the image
is \fg. The significance level is given by the color scale, with the
contours start at $3\sigma$, and following steps of $1\sigma$. The position of \fg\ (magenta) as well as its updated center following the
2FGL {\it Fermi}-LAT Catalog (blue)\footnotemark  are shown, while the cross represent the counterparts identified using XRT (from Corbet et al. 2011). }
\label{mos-1}
\end{figure*}

\footnotetext { See the website: \url{http://fermi.gsfc.nasa.gov/ssc/data/access/lat/2yr_catalog/} for more information on 2FGL {\it Fermi}-LAT Catalog.}

An \textit{INTEGRAL} detection of 1FGL J1018.6-5856 is derived from combining all the ISGRI data, with a 
significance level of 5.4$\sigma$ and an average intensity of 0.074 counts s$^{-1}$ in the 18--40 keV band. 
1FGL J1018.6-5856 is also consistently detected  in the 18--60 keV;  its map significance is reduced only to a fluctuation at higher energies. 
We only  obtain 1.63$\sigma$  in the 40--100 keV.
The significance and flux measurement of 1FGL J1018.6-5856 are obtained from the pixel comprising the most prominent position  determined by {\it Swift}-XRT (Corbet et al. 2011), which is within {\it Fermi}-LAT 1FGL and 2FGL error circle.
The pixel size of \textit{INTEGRAL} is about 4.93 arcmin $\times$ 4.93 arcmin, whereas the uncertainty in position of \textit{Fermi}, i.e., the 95\% confidence radius is $\sim$1.76 arcmin and is thus fully included in the pixel used for our significance estimation.
However, we would like to notice that the source significance should be taken with care.
The pixel significance  distribution of each 18-40 keV map has
a Gaussian shape with a  width parameter, instead of 1 as expected from a purely statistical dominated image,  of  about 1.1, which may due to an unknown systematic error. Typically, 
 systematic noise is produced from being close to the very bright sources and or very crowded regions. 
 As more maps from different observations are
 combined, the width parameter increases and finally reaches up to 1.4 for all data considered. 
 Thus, we find that there is a non-statistical
noise in the image that effectively reduces the significance to 5.4/1.4, or about 4$\sigma$. 
Given the spatial coincidence of the source with \fg\  we can claim a detection at  99.9937 \%, 
or alternatively, that the probability of a chance coincidence is of the order $\sim$ $6.3 \times 10^{-5}$.

If we assume (although there is no a prior reason to, and we caveat that this is just done for estimation to an order of magnitude) that the spectral shape of  \fg\ is  same with that of the Crab nebula in the corresponding energy range (a power-law with $\Gamma\sim$2.14),  a flux of \fg\ is derived as  $2.25\times10^{-11}$ erg cm$^{-2}$  s$^{-1}$, or about 0.35 mCrab in flux units. If we instead assume that the spectrum of \fg\ is similar to that of \ls\ at inferior conjunction passage as given by Hoffmann et al. (2009), we would obtain \fg\  flux of  about 
$0.75\times10^{-11}$ erg cm$^{-2}$  s$^{-1}$, or  a factor of 0.23  of the LS 5039 flux.   The neighboring source GRO J1008-57, 1.36$^\circ$ away from \fg\, is detected with a significance of $54.6\sigma$ and with an intensity of 4.2 mCrab. It  likely constitutes  a problem for the analysis of \fg\ with non-imaging detectors.

\begin{table} 
\centering
\scriptsize
\caption{Exposure times and obtained significance 
for the whole \textit{INTEGRAL}   observation (all phase bins) and separated in orbital bins.  }
\begin{tabular}{ccccc }
\hline
phase  &  flux & significance    &   exposure & width of combined sig. \\
  & (counts s$^{-1}$) & (in 18--40 keV) &   (Ms) & map distribution\\
\hline

0.0--1.0      &  0.074$\pm$0.014&           5.4                &              5.78  & 1.39\\
\hline
0.0--0.2   &    -0.0073$\pm$0.032            &               --         &    1.16                       &          1.36 \\
%ldots=-0.00725265+/-0.03181
0.2--0.4   &    0.094$\pm$0.030   &            3.12     &          1.13            &                     1.32 \\
0.4--0.6   &    0.154$\pm$0.031    &            4.89      &        1.13        &                         1.37 \\
0.6--0.8   &    0.050$\pm$0.032  &             1.54       &       1.05    &                             1.30 \\
0.8--1.0   &    0.061$\pm$0.028  &              2.15         &     1.31   &                               1.36 \\
\hline
\end{tabular}
\label{expo-hard}
\vspace{0.1cm}
\end{table}

Since \fg\ is much weaker than other neighboring sources in its crowded sky region, it is impossible for OSA 9.0 to derive a lightcurve directly out of the standard reduction procedure. However, one can combine the images from different science windows based on orbital phase-bins and, via inferring the flux from each of the combined image, produce an orbital lightcurve manually. Thus, to see how \fg\ varies along its 16.58 days orbit, we divide the \textit{INTEGRAL} data into phase bins. The reference time at phase zero is set to the peak flux that observed by {\it Fermi-LAT} at $T_{max}={\rm MJD} \; 55403.3 \pm n \times 16.58 $ days, as reported at GeV energies in Corbet et al. (2011).  We show the  \textit{INTEGRAL} results on 1FGL J1018.6-5856 for each phase bin in Table \ref{expo-hard}, where the  flux, the corresponding   exposure  and the significance are provided.  Again, given that the pixel distribution  has a Gaussian  width of about 1.34 in each of the images (see Table \ref{expo-hard}), we conservatively consider that the significance should be lowered by this factor.
Hinted from Table 1 is a trend of having an anti-correlation between the hard X-ray emission and the {\it Fermi}-LAT periodicity. 
Actually, we find that a significant portion of the signal comes only from a 0.2-wide phase region, at phases 0.4--0.6 (where even the corrected significance amounts to about 90\% of the signal found in total). However, the scarcity of counts makes it difficult to have a definitive proof of the variability: a constant fit to the count rate yields a reduced $\chi^2$ of 14.09/4, suggesting that the significance of variability is at 99.64\%, or only 2.7$\sigma$ level.

Though it is impossible to derive a lightcurve from standard reduction procedure, we could
read pixels corresponding to \fg' position and derived a light curve 
manually. To search for a periodic signal in the light curve data, we
used the Lomb--Scargle periodogram method. Power spectrum is generated for the 
light curve using the PERIOD subroutine (Press \& Rybicki 1989). No significant signal is seen at 
an orbit period of 16.58 days beyond  90\% white--noise confidence level. This is consistent with  having only 
an overall  weak detection of the source in our imaging analysis and the orbit modulation is not apparent in lightcurve.

\section{Concluding remarks}

We have carried out an analysis of all   \textit{INTEGRAL} data available to 1FGL J1018.6-5856, a 
new gamma-ray binary with a period of $\sim$ 16.5 days discovered blindly by means of a power spectrum analysis
of {\it Fermi}-LAT detection (100 MeV--200 GeV). The total effective exposure   extracted on the source from the archive amounts to 5.78 Ms, and leads to a credible detection of the source at hard X-rays (18--40 keV). The count rate is however very low,  preventing from   extracting strong conclusions about orbital variability, although it is hinted with an anti-correlation with the gamma-ray emission detected by {\it Fermi}-LAT in 100 MeV -- 200 GeV. In fact, 1FGL J1018.6-5856 seems to significantly show up only during a small part of its orbital evolution, far from the gamma-ray maximum at 100 MeV--200 GeV. This would represent a more marked distinction with respect to  what was obtained in the case of \lsi\ which presents a displacement of the maximum between  hard X-rays (by \textit{INTEGRAL}) and gamma-rays (by {\it Fermi}-LAT and MAGIC), without being completely anti-correlated (Zhang et al. 2010). Instead this is in line with the results for LS 5039 (see, e.g., Hoffman et al. 2009) where the hard X-ray emission as measured with \textit{INTEGRAL} 
is correlated with the TeV emission measured with HESS (Aharonian et al. 2006), and thus it is fully anti-correlated with the GeV emission as measured by {\it Fermi}-LAT (Abdo et al. 2009). The case of 1FGL J1018.6-5856 appears as another incarnation of this behavior, albeit with the conservative caveat of yet a low significance for an strong claim ---the maximum count rate of 1FGL J1018.6-5856 is about five times lower than that of LS 5039---,
emphasizing a possible physical similarity of the two sources. TeV observations at what appears to be the maximum of the hard X-ray lightcurve are thus encouraged.

\acknowledgements

We acknowledge support from the grants AYA2009-07391 and SGR2009-811, as
well as the Formosa Program TW2010005. NR is supported by a Ramon y Cajal Fellowship. 
This work was also subsidized by the National Natural Science Foundation of China via NSFC-10325313,10521001,10733010,11073021 and 10821061, the CAS key Project KJCX2-YW-T03,
and 973 program 2009CB824800

\end{document}